\documentclass{PoS}

\title{The LOFAR Known Pulsar Data Pipeline }

\ShortTitle{The LOFAR Known Pulsar Data Pipeline }

\author{\speaker{Anastasia Alexov}%
        \thanks{Thank you to the LOFAR Pulsar Working Group and the LOFAR
       Transients Key Science Project}\\
       University of Amsterdam (UvA), Amsterdam, The Netherlands\\
       E-mail: \email{a.alexov@uva.nl}}

\author{Jason W. T. Hessels\\
        Netherlands Institute for Radio Astronomy (ASTRON), Dwingeloo, The Netherlands\\
        University of Amsterdam (UvA), Amsterdam, The Netherlands\\
        }

\author{Jan David Mol\\
        Netherlands Institute for Radio Astronomy (ASTRON), Dwingeloo, The Netherlands\\
        }

\author{Ben Stappers\\
        The University of Manchester, Manchester, England\\
        }

\author{Joeri van Leeuwen\\
        Netherlands Institute for Radio Astronomy (ASTRON), Dwingeloo, The Netherlands\\
        University of Amsterdam (UvA), Amsterdam, The Netherlands\\
        }

\abstract{
Transient radio phenomena and pulsars are one of six LOFAR Key Science Projects (KSPs).  As part of the Transients KSP, the Pulsar Working Group (PWG) has been developing the LOFAR Pulsar Data Pipelines to both study known pulsars as well as search for new ones.  The pipelines are being developed for the Blue Gene/P (BG/P) supercomputer and a large Linux cluster in order to utilize enormous amounts of computational capabilities (50Tflops) to process data streams of up to 23TB/hour.  The LOFAR pipeline output will be using the Hierarchical Data Format 5 (HDF5) to efficiently store large amounts of numerical data, and to manage complex data encompassing a variety of data types, across distributed storage and processing architectures.  We present the LOFAR Known Pulsar Data Pipeline overview, the pulsar beam-formed data format, the status of the pipeline processing as well as our future plans for developing the LOFAR Pulsar Search Pipeline.  These LOFAR pipelines and software tools are being developed as the next generation toolset for pulsar processing in Radio Astronomy.
}

\FullConference{ISKAF2010 Science Meeting - ISKAF2010\\
		June 10-14, 2010\\
		Assen, the Netherlands}

\begin{document}

\section{LOFAR Observing}

LOFAR operates in two frequency ranges: 10-80\,MHz, using
the Low Band Antennas (LBA), and 100-240\,MHz, using the High Band
Antennas (HBA).  Close to 6000 antennas are distributed among 40 stations in
the Netherlands, with several additional stations completed or
planned throughout the rest of Europe.  An observation can use subsets
of any or all of the stations available.  Radio interferometric imaging at 
$\sim$1-second integration times is the main observing mode.
However the high-time-resolution beam-formed observing modes will also be widely used for transient science.
In this non-imaging mode, spatial resolution is
traded for time resolution and
time sampling can be as short as 5\,$\mu$s. When multiple fields of view are synthesized
simultaenously, the data rate can be as large as 23TB/hr.
  
The Pulsar Working Group is in the process of creating automated
pipeline processing software for studying and finding pulsars.
This Pulsar Data Pipeline effort is the basis for other high-timing-resolution
pipelines using LOFAR to study, e.g., the Sun and planets.

\section{LOFAR Transient Science \& Requirements on Processing}

The processing of time-series (Beam-Formed) data has many scientific applications.  The
following areas of study set the requirements on processing for each
of the envisioned sub-modes: (1) \textit{Known pulsars}: channelization, Stokes parameters, dedispersion, folding, radio frequency interference (RFI) excision;  (2) \textit{Pulsar/fast transient survey}: channelization, dedispersion [1000s of trial dispersion measures (DMs)], RFI excision, searching; (3) \textit{Planets, Sun, flare stars}: channelization, RFI excision, dynamic spectra.

Starting with LOFAR observations of known pulsars, the group has
designed and implemented a pipeline for processing LOFAR time-series
data, checking for pulsar detection, and analyzing the results~\cite{Stappers11}.  The
pipeline implements new and existing software tool sets for data
processing, described in the sections below.

\section{Present LOFAR Beam-Formed (BF) Data Flow \& Processing}

\begin{figure}
\includegraphics[angle=90, width=.80\textwidth]{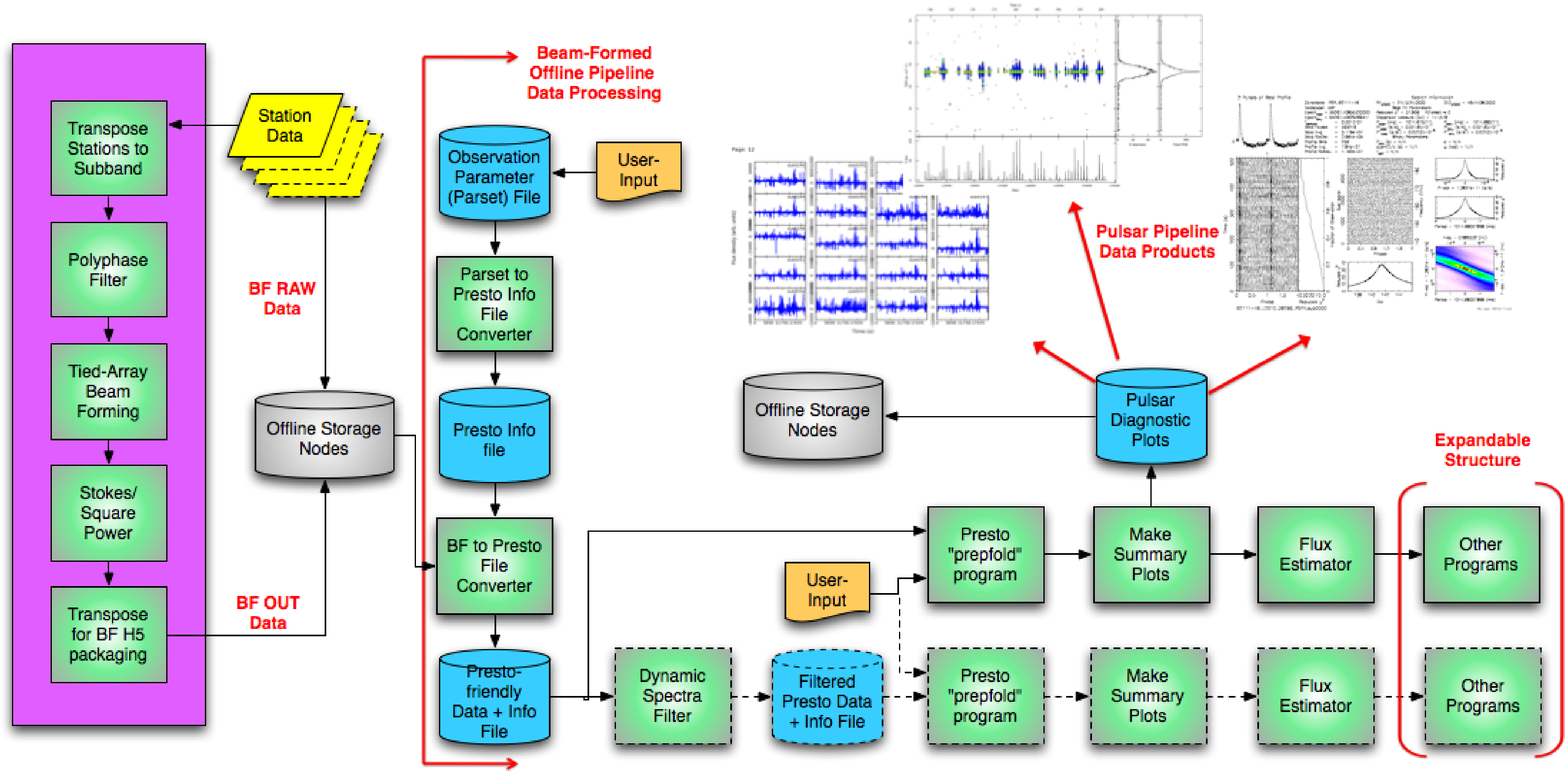}
\caption{LOFAR Beam-Formed Data Flow: online BG/P processing steps are shown on the left and offline processing steps in the center and right.  Presto is a community standard software suite for pulsar data analysis,
including dedispersion, folding, and search algorithms.} \label{known_pulsar_pipeline_overview}
\end{figure}

The Beam-Formed data flow and processing are split into two broad, consecutive segments:
``online" processing, which is done by a Blue Gene P (BG/P) supercomputer to
streaming data from the LOFAR stations, and ``offline" processing
which is further processing of these data on the LOFAR offline
cluster.  The online processing is chiefly responsible for combining
data from multiple stations into one or multiple beams, as
well as optionally producing correlations for imaging.  The
offline processing is more science-specific;  it includes, for example
dedispersion and folding of the data at a known pulsar period; it is 
even more flexible than the ``online" processing system.  Figure~\ref{known_pulsar_pipeline_overview} 
shows the overall Beam-Formed data flow for a pulsar observation.   The
``online" and ``offline" division is marked by the vertical red line.

Given that the processing of the raw station data is done entirely in
software, the pipeline is quite flexible and extendable.  Here we
describe the current implementation.  Many
of the various online Pulsar Pipeline steps are
common to other LOFAR observation modes.

\subsection{Online: Tied-Array Beam Pipeline}

The station signals are fed to a Blue Gene P (BG/P)
supercomputer in Groningen which can both combine the signals into
coherent or incoherent array beams~\cite{vL&S2010}, or correlate them for imaging.  
These steps are on the left hand side of Figure~\ref{known_pulsar_pipeline_overview},
within the purple frame.  In brief, the steps of the online Tied-Array Beam (TAB) Pipeline
are: 
(1) \textit{First Transpose}: for efficiency, data are transposed between stations and frequency subbands; 
(2) \textit{Second Polyphase Filter}: an optional number of channels ($16-256$) is created within each 156/195-kHz subband via a polyphase filter (PPF); 
(3) \textit{Beam-Forming}: coherent and/or incoherent beam-forming between multiple stations (optional); 
(4) \textit{Stokes Parameters, Downsampling, and Re-binning}: station signals from the X and Y
polarizations of the antennas are used to derive
the 4 Stokes parameters, I, Q, U, and V, or, to
reduce the data rate by a factor of 4, one can optionally output only
Stokes I, the total intensity.  

\subsection{Offline: Pipeline Processing}

With the station data now (optionally) combined into
incoherent/coherent array beams, the goal of the offline processing is
to apply further data-reduction.  These steps are shown in the center and right hand side of 
Figure~\ref{known_pulsar_pipeline_overview} as modules of the
``offline" Known Pulsar Pipeline.  In general, this pipeline performs 
RFI mitigation, dedispersion, and
folding.

Offline processing happens on the LOFAR Offline Cluster, also in Groningen.  The
cluster is divided into subclusters, each of which contains 120TB of disk storage
and 9 compute nodes, with 2 quad-core processors each.  In general, two subclusters are
currently used for processing pulsar observations, which gives us over 
200TB of disk space and 144 cores
(split between 18 machines).

\subsubsection{Known Pulsar Pipeline}

For observations of known pulsars, the primary offline processing steps are 
dedispersion at the pulsar dispersion measure (DM) and folding using a pulsar 
ephemeris.  These steps are automated in the Known Pulsar Pipeline that 
predominantly uses tools from the PRESTO\footnote{http://www.cv.nrao.edu/\~{}sransom/presto/}
\& TEMPO\footnote{http://www.atnf.csiro.au/research/pulsar/tempo/} software suites.  Figure~\ref{known_pulsar_pipeline_overview} (center and right) shows all the steps in the Offline Known Pulsar Pipeline;  several steps are preparatory and data-format-related, followed by the use of science tools. 
Several pipeline processing modes are available: (1) \textit{Incoherent Stokes}: station data are summed incoherently; (2) \textit{Coherent Stokes}:  station data are summed coherently; (3) \textit{Fly's Eye}: station data are recorded separately for the purposes of system tests or to form a very large field of view.

RFI checking is also performed for long-term commissioning statistics, and for future RFI-mitigation.  
Pulsar diagnostic plots are produced (see top right of Figure~\ref{known_pulsar_pipeline_overview}) in order to quickly discern whether a pulsar was detected in the data.  A typical 1-hour observation can be processed in 20 minutes using 8-cores on one machine, making it easy to reduce the data in realtime.  

Since pipeline 
processing is modular, we can easily change the Known Pulsar Pipeline to use 
other pulsar tools if need be;  we have the capability of using several file 
formats so that all existing pulsar tools can interface with
the LOFAR data format.  Several major upgrades (see Sections~\ref{Pipeline Framework} and~\ref{Data Format}) are being worked on for the Known Pulsar Pipeline, which will also be applicable for other LOFAR pipelines, such as the Pulsar Search Pipeline.  

\subsubsection{Pulsar Search Pipeline}

We are in the process of testing a proto-type LOFAR Pulsar Search
Pipeline, which uses several of the standard search tools from the
PRESTO software suite.  This proto-type pipeline is
blindly detecting known pulsars in LOFAR data;  it can
manage RFI easily; and, it can run at close to real time if we use
9 compute nodes (72 cores in total) and search only for "slow" pulsars (P\_spin $\gtrsim$ 50ms).

\subsubsection{Pipeline Framework}
\label{Pipeline Framework}

Like other LOFAR processing pipelines, the Pulsar Pipelines are being built to
run within a generic pipeline framework.  This framework takes care of
distributing the processing over the LOFAR offline cluster and its nodes,
and provides appropriate logging and error checking\cite{Swinbank10}.  The Pipeline Framework manages large 
parallel jobs in a more natural way, so that active observations and 
post-processing do not clash on the cluster.  There is significant speedup 
by using the framework, which intelligently distributes the 
processing to all 9 machines within a subcluster, with 8 cores each.  In the production
system, the Pulsar Pipelines will start processing automatically after
an observation.

\subsubsection{Data Format}
\label{Data Format}

The data from the BG/P are
currently written as almost header-less raw binary files and then converted to
PRESTO-friendly format for pipeline processing.  We can also convert data to
SIGPROC\footnote{http://sigproc.sourceforge.net/} "filterbank" and 
"PSRFITS"~\cite{vanStraten:2009ct} formats, thereby 
expanding the repertoire of pulsar tools we can use in the pipelines. 
The LOFAR Pulsar data format will soon be based on an implementation of the Hierarchical Data Format
5 (HDF5)\footnote{HDF home page: http://www.hdfgroup.org/}.  The HDF5 
file format was chosen because of its flexibility, its versatile data model, its ability to store very large, complex datasets spread over many separate physical devices.  
Anderson et al. (2010) discusses in depth the benefits of using HDF5 for LOFAR data storage,as well as the HDF5-tools which can be used for plotting and visualization of the data. 
The LOFAR Beam-Formed HDF5 data format is
described in detail in the LOFAR Interface Control Document 
(ICD) \#3~\cite{Alexov10}.

\section{Software and Access}
\label{Software and Access}

As much as possible, the Pulsar Pipeline tries to use the well-tested,
commonly used, open software packages available in the pulsar
community (e.g. PRESTO, 
PSRCHIVE\footnote{http://sourceforge.net/projects/psrchive/} v13.0, 
SIGPROC v4.3, 
TEMPO v1.1
and Python bindings to PGPLOT, called 
ppgplot\footnote{http://www.astro.rug.nl/~breddels/python/ppgplot/} v1.1).  
These have been
incorporated into the general LOFAR User Software 
(LUS)\footnote{http://usg.lofar.org}, which uses 
Cmake\footnote{http://www.cmake.org/}, a cross-platform, open-source 
build system, to automate the installation of these packages on a
variety of platforms (e.g. Ubuntu Linux and Mac OS X).

The current LOFAR Known Pulsar Pipeline uses PRESTO modules for folding data and viewing the results.
PRESTO is a pulsar software tool suite
that has been used within the pulsar community for almost 10
years.   Several community-driven pulsar analysis tools complementary to PRESTO
are being considered for LOFAR pulsar data pipeline processing - PSRCHIVE
and SIGPROC.  Both tool suites have had extensive use in the
pulsar community and can be integrated as pipeline and analysis tools for LOFAR BF data.
Using ``off-the-shelf" software like PRESTO, SIGPROC and PSRCHIVE saves development time and also significant 
test and commissioning time, and, any software improvements to these tools
will be fed back into the pulsar community.

\section{Beam-Formed Data Processing Enhancements}
\label{Beam-Formed Data Processing Enhancements}

In the coming year, a number of enhancements are anticipated for 
LOFAR Beam-Formed data processing:
integrate the Known Pulsar Pipeline into the LOFAR Pipeline Framework 
to gain distributed processing, parallelization, logging and automated-kickoff 
after an observation; add coherent dedispersion to the Known Pulsar Pipeline;  
finalize the HDF5-writer to output BF data conforming to the 
LOFAR BF ICD; integrate pipeline tool suite with 
HDF5 data format;  design Pulsar ``searching" mode pipelines to reduce 
extreme data rates to few GB; automatically archive BF data products;  
implement ``Piggyback" simultaneous observing with the LOFAR Million Source 
Shallow Survey (MSSS);  add polarimetry and single pulse 
pulse data processing to the pipeline.

The LOFAR observatory is in the process of executing commissioning 
proposals, including pulsar observations.  The current LOFAR Known Pulsar Pipeline
has already been used to automatically process and detect
more than 100 known pulsars~\cite{Hessels10}.  The overall data-taking/processing/archiving system is being worked to
accommodate pulsar and other LOFAR observations.  The LOFAR Known and Search
Pulsar Pipelines are anticipated to be fully automated within this system in the
coming year, with many of the enhancement features listed above. This automation, 
parallel processing and distributed file-systems will help solve the enormous challenge of
reducing 23TB/hr of data into several tens of GB of highly processed data:
  e.g., a pulsar ``survey" observation is reduced to the
parameters of the (new) candidate pulsars that were found!

\end{document}